\documentclass[twocolumn,showpacs,preprintnumbers,amsmath,amssymb]{revtex4}

\usepackage{graphicx}
\usepackage{dcolumn}
\usepackage{bm}

\begin{document}

\title{Control momentum entanglement with atomic spontaneously generated coherence}

\author{Rui Guo}
\author{Hong Guo}\thanks{Author to whom correspondence should be
addressed. E-mail: hongguo@pku.edu.cn, phone: +86-10-6275-7035, Fax:
+86-10-6275-3208.}

\affiliation{CREAM Group, School of Electronics Engineering $\&$
Computer Science, Peking University, Beijing
100871, P. R. China\\}%

\date{\today}

\begin{abstract}
With atomic spontaneously generated coherence (SGC), we propose a
novel scheme to coherently control the atom--photon momentum
entanglement through atomic internal coherence. A novel phenomena
of ``phase entanglement in momentum'' is proposed, and we found,
under certain conditions, that super--high degree of momentum
entanglement can be produced with this scheme.
\end{abstract}

.

\maketitle

\section{Introduction}

Entanglement with continuous variable attracts many attentions for
its fundamental importance in quantum nonlocality \cite{EPR} and
quantum information science and technology \cite{rmp}. As a
physical realization, the continuous momentum entanglement between
atom and photon has been extensively studied in recent years
\cite{Singlephoton,3-D spontaneous,scattering,GR,exp}. In the
process of spontaneous emission \cite{Singlephoton,3-D
spontaneous}, the momentum conservation will induce the
atom--photon entanglement, with its degree inversely proportional
to the linewidth of the emission. Therefore, by squeezing the
effective transition linewidth, super--high degree of momentum
entanglement may be produced \cite{scattering,GR}. With this
entanglement, further, it is possible to realize the best
localized single--photon wavefunction even in free space
\cite{Singlephoton}.

In this paper, we propose a novel scheme to coherently control and
enhance the momentum entanglement between single atom and photon.
We found, for the atomic system with spontaneously generated
coherence (SGC) \cite{SGC,atomic coherence contr}, that the
interference between photons emitted along different quantum
pathways could enhance the momentum entanglement significantly.
Due to SGC, the degree of entanglement is determined by the
intensity of the interference in the emission process, and can be
effectively controlled by the atomic coherence between its
internal states. Under this new mechanism of entanglement
enhancement, the entangled system could exhibit novel feature of
``phase entanglement'' in the momentum space, which does not exist
in the previous schemes without interference
\cite{Singlephoton,3-D spontaneous,scattering,GR,exp}; and the
degree of entanglement is found to be ``abnormally'' proportional
to the atomic linewidth. Moreover, by effectively squeezing the
separation of the upper levels, it is possible to produce
super--high degree of momentum entanglement for the atom--photon
system with this scheme \cite{GR-disentanglement}.

\begin{figure}
\centering
\includegraphics[height=2.8cm]{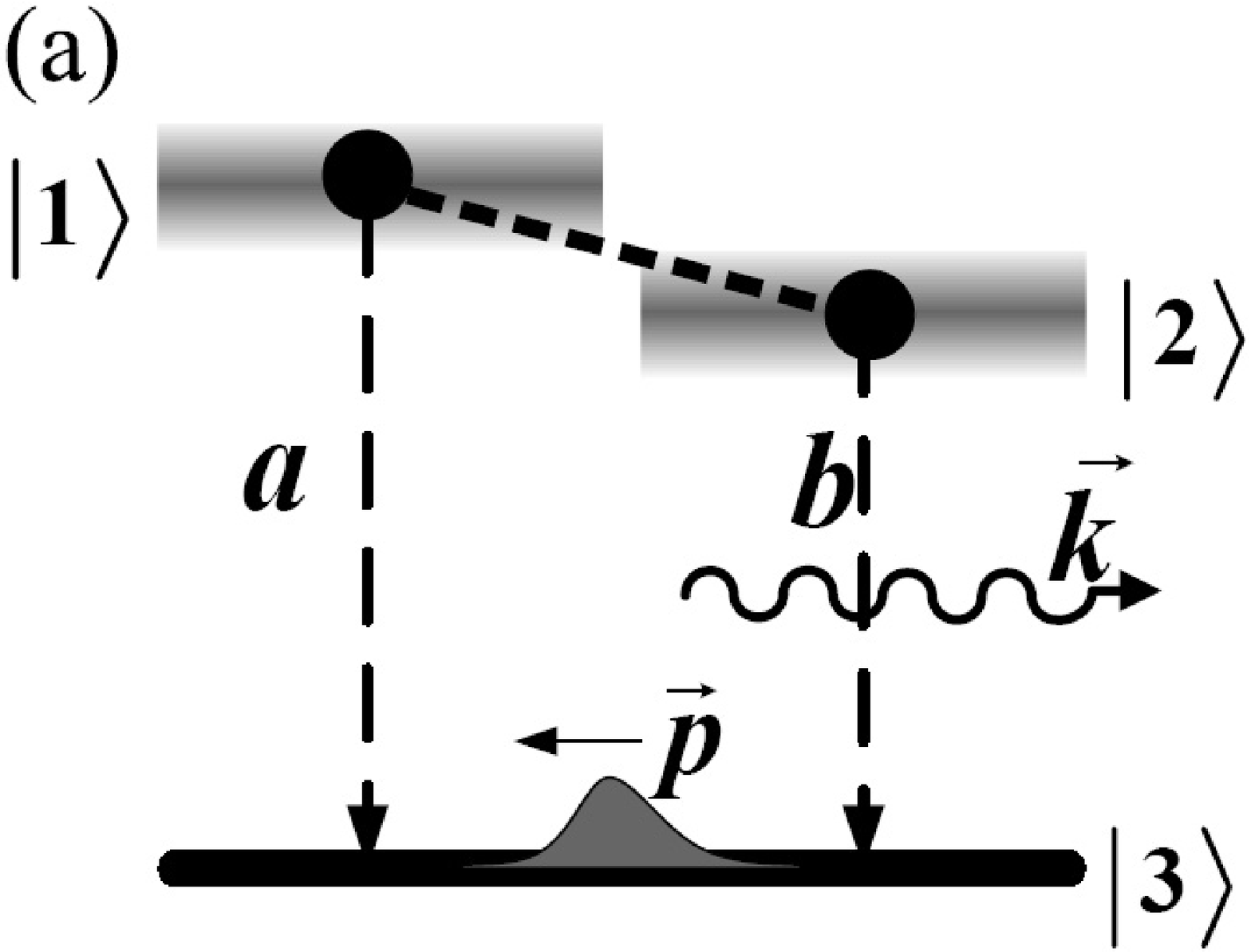}\ \ \ \ \ \
\includegraphics[height=2.8cm]{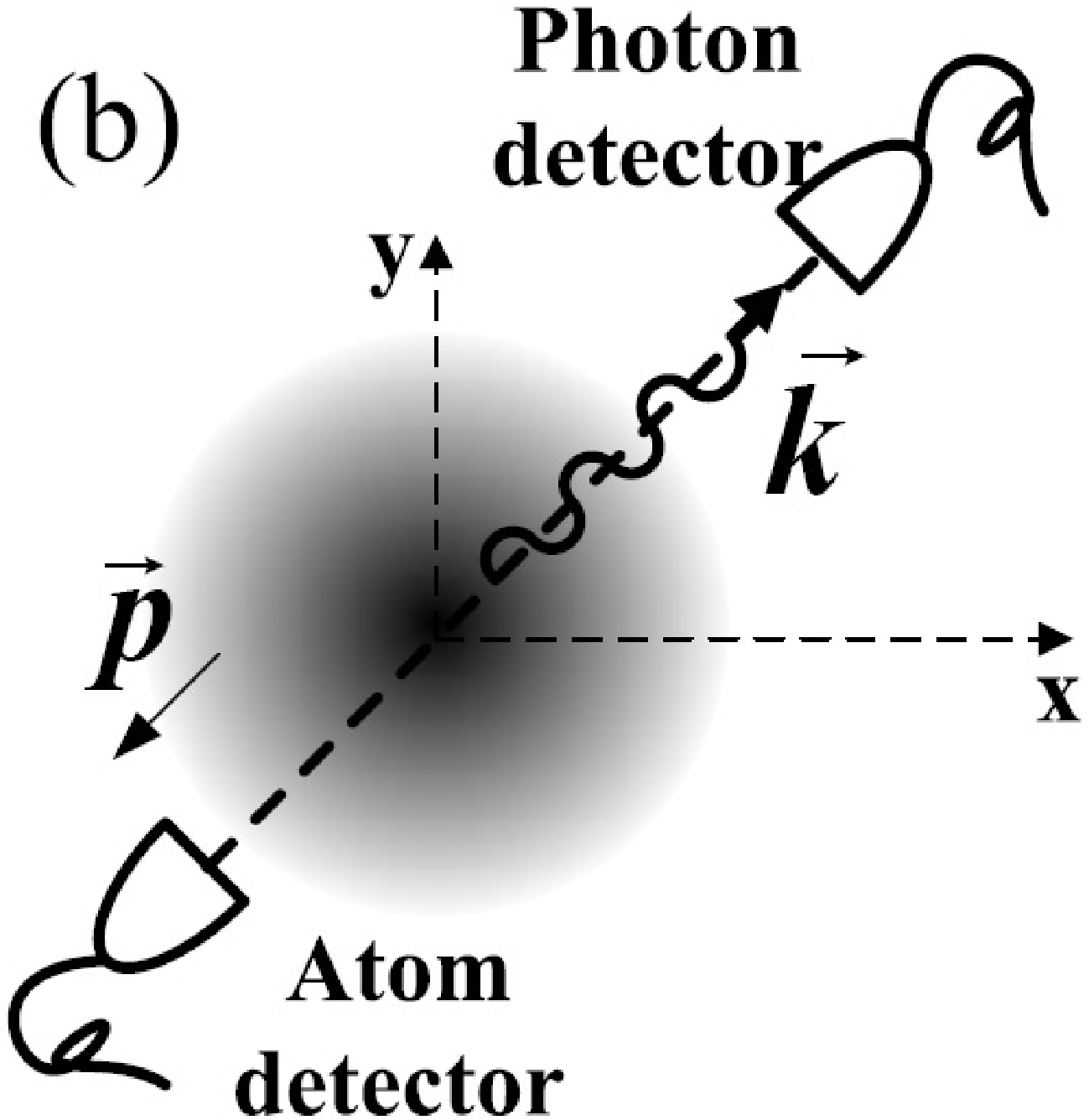}
\caption{(a) The atom has two closely--lying upper levels to
provide different quantum pathways for the spontaneous emission.
The momentum conservation makes the emitted photon entangled with
the recoiled atom. (b) Schematic diagram for the detections. The
two detectors are fixed in one dimension as in the reported
experiments \cite{exp}.}
\end{figure}

\section{Theoretical model}

As shown in Fig. 1 (a), the atom with two nearly--degenerate upper
levels has two transition pathways (denoted by ``a'' and ``b''
respectively) to induce the momentum entanglement with the emitted
photon. To have strong interference between the two transitions,
we assume that the dipoles $\mu_{a,b}$ for the transitions are
parallel with each other \cite{SGC}. Then, the Hamiltonian under
the rotating wave approximation (RWA) can be written as:
\begin{eqnarray}
\hat{H}&=&\frac{(\hbar \hat{\vec{p}})^{2}}{2m}+\sum_{\vec{k}}\hbar
\omega_{\vec{k}}\hat{a}^{\dag}_{\vec{k}}\hat{a}_{\vec{k}}+\hbar
\omega_{a}\hat{\sigma}_{11}+\hbar \omega_{b}\hat{\sigma}_{22}\\
\nonumber
 &+&\hbar\sum_{\vec{k}}\left[g_{a}(\vec{k})\hat{\sigma}_{31}\hat{a}^{\dag}_{\vec{k}}e^{-i\vec{k}\cdot\vec{r}}
+g_{b}(\vec{k})\hat{\sigma}_{32}\hat{a}^{\dag}_{\vec{k}}e^{-i\vec{k}\cdot\vec{r}}+{\rm
H.c.} \right],
\end{eqnarray}
where $\hbar \hat{\vec{p}}$ and $\vec{r}$ denote atomic
center--of--mass momentum and position operators,
$\hat{\sigma}_{ij}$ the atomic operator $|i\rangle \langle j|$
($i,j=1,2,3$), and $\hat{a}_{\vec{k}}$
($\hat{a}^{\dag}_{\vec{k}}$) is the annihilation (creation)
operator for the $k$th vacuum mode with wave vector $\vec{k}$ and
frequency $\omega_{\vec{k}}\equiv ck$. $g_{a,b}(\vec{k})$ are the
coupling coefficients for the transitions ``a'' and ``b'', where
we use $\vec{k}$ to denote both the momentum and polarization of
the vacuum mode for simplicity.

With the spontaneous emission, the momentum conservation will make
the emitted photon entangled with the recoiled atom in momentum.
It is convenient to depict this entangling process in the
Schr\"{o}dinger picture, and expand the photon--atom state as:
\begin{eqnarray}
\nonumber |\psi\rangle&=&\sum_{\vec{q}}C_{1}(\vec{q},t)
|\vec{q},0,1\rangle+\sum_{\vec{q}}C_{2}(\vec{q},t)
|\vec{q},0,2\rangle\\
&+&
\sum_{\vec{q},\vec{k}}C_{3}(\vec{q},\vec{k},t)|\vec{q},1_{\vec{k}},3\rangle\
,
\end{eqnarray}
where the arguments in the kets denote, respectively, the wave
vector of the atom, the photon, and the atomic internal states.

With the transformation
\begin{eqnarray}
C_{1,2}(\vec{q},t)&=&\exp \left [-i \left ( \frac{\hbar
q^{2}}{2m}+\omega_{a}
\right )t \right ]\cdot A_{1,2}(\vec{q},t)\ ,\\
C_{3}(\vec{q},\vec{k},t)&=&\exp \left [-i \left ( \frac{\hbar
q^{2}}{2m}+c k \right )t \right ]\cdot B(\vec{q},\vec{k},t)\ ,
\end{eqnarray}
and Weisskopf--Wigner approximation, we yield the dynamic
equations from the Schr\"{o}dinger equation:
\begin{eqnarray}
\frac{{\rm d}A_{1,2}(\vec{q},t)}{{\rm
d}t}&=&-\frac{\gamma_{a,b}}{2}A_{1,2}(\vec{q},t)\nonumber \\
&-&\frac{\varepsilon
\sqrt{\gamma_{a}\gamma_{b}}}{2}A_{2,1}(\vec{q},t)e^{\pm i(\omega_{a}-\omega_{b})t},\\
i\frac{{\rm d}B(\vec{q},\vec{k})}{{\rm d}t}&=&
g_{a}(\vec{k})e^{i[-\hbar(2\vec{q}+\vec{k})\cdot
\vec{k}/2m+ck-\omega_{a}]t}A_{1}(\vec{q}+\vec{k}) \nonumber \\
&+& g_{b}(\vec{k})e^{i[-\hbar(2\vec{q}+\vec{k})\cdot
\vec{k}/2m+ck-\omega_{b}]t}A_{2}(\vec{q}+\vec{k}) \ , \nonumber\\
\end{eqnarray}
where the nonrelativistic approximation ($\hbar \vec{q},\ \hbar
\vec{k} \ll mc $) and the relation $\omega_{12}\equiv
\omega_{a}-\omega_{b}\ll \omega_{a},\omega_{b}$ are used.
$\gamma_{a,b}=2\pi \sum_{\vec{k}}|g_{a,b}(\vec{k})|^{2}\delta
[\hbar(2\vec{q}-\vec{k})\cdot\vec{k}/2m +\omega_{a,b}-ck]$ are the
linewidthes for the two transitions, and $\varepsilon\equiv
\vec{\mu}_{a}\cdot\vec{\mu}_{b}/|\vec{\mu}_{a}|\cdot|\vec{\mu}_{b}|=1$
as we have assumed previously. Suppose the atom is initially
prepared to a superposed state $A_{10}|1\rangle+A_{20}|2\rangle$
and has a Gaussian wavepacket as $G(\vec{q})\propto
e^{-(\vec{q}/\delta q)^{2}}$, and the detections are restricted in
one dimension as in Fig. 1 (b), the one--dimensional steady
solutions for Eqs. (5) and (6) yield:
\begin{eqnarray}
&&A_{1}(q,t\rightarrow \infty )=A_{2}(q,t\rightarrow\infty)=0 \
,\\
&& B(q,k,t\rightarrow\infty)=-i\chi_{0}\exp[-(\Delta
q/\eta)^{2}]\times \\
&& \left[\frac{C_{1}(2g_{b}s_{1}/\varepsilon
\sqrt{\gamma_{a}\gamma_{b}}-g_{a})}{i(\Delta q+\Delta
k)+(\frac{s_{1}}{\gamma_{a}}-\frac{1}{2})} +
\frac{C_{2}(2g_{b}s_{2}/\varepsilon
\sqrt{\gamma_{a}\gamma_{b}}-g_{a})}{i(\Delta q+\Delta
k)+(\frac{s_{2}}{\gamma_{a}}-\frac{1}{2})} \right] , \nonumber
\end{eqnarray}
where the parameters are defined as:
\begin{eqnarray}
&& s_{1,2}\equiv\frac{1}{2}(\lambda
\pm\sqrt{\lambda^{2}+\varepsilon^{2}\gamma_{a}\gamma_{b}})\ ,\\
&& \lambda\equiv\frac{1}{2}(\gamma_{a}-\gamma_{b})+i\omega_{12}\ ,\\
&& C_{1,2}\equiv
\pm\frac{s_{2,1}A_{10}+\frac{1}{2}\varepsilon\sqrt{\gamma_{a}\gamma_{b}}A_{20}}{s_{2}-s_{1}}\
,\\
&& \eta\equiv\frac{\delta q \hbar k_{0}}{m \gamma_{a}}\ ,  \ \ \ \
\ \ \ k_{0}\equiv\frac{\omega_{a}}{c} \ ,
\end{eqnarray}
and the effective wave vectors are defined by:
\begin{eqnarray}
\Delta k &\equiv&\frac{k-k_{0}}{\gamma_{a}/c}\ ,  \ \ \ \ \ \Delta
q \equiv\frac{\hbar k_{0}}{m \gamma_{a}}(q-k_{0})\ ,
\end{eqnarray}
where $\chi_{0}$ is the normalized coefficient.

\section{amplitude entanglement in momentum}

\begin{figure}
\centering
\includegraphics[height=4cm]{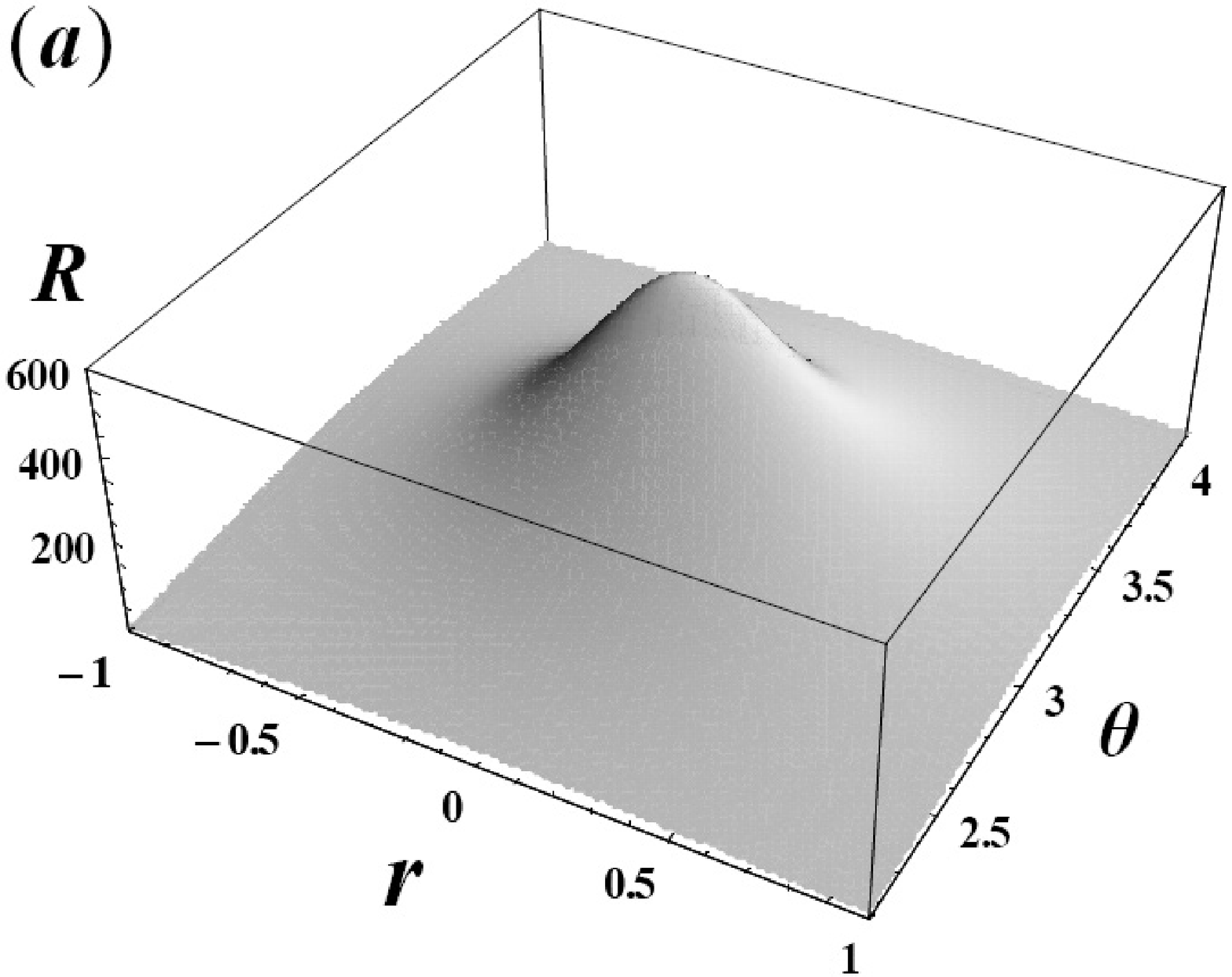}\includegraphics[height=4cm]{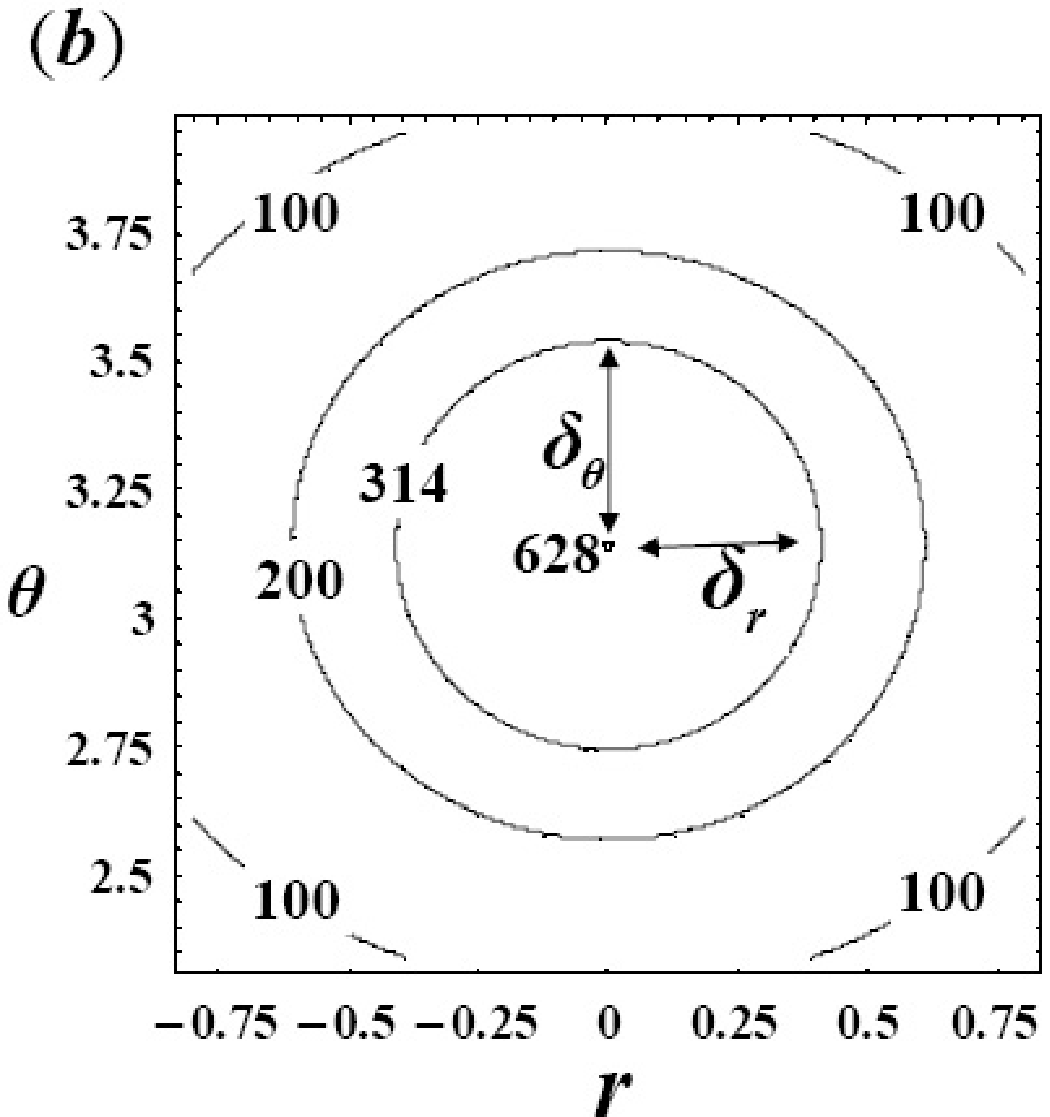}
\caption{(a) The ``amplitude entanglement'' $R$ ratio is plotted
in dependence on the atomic coherence $r$ and $\theta$ with
$\delta=0.02$, $\eta=0.1$. (b) The contour plot of figure (a). The
circular contours indicates the symmetric roles played by $r$ and
$\theta$ in controlling the $R$ ratio. The FWHM is denoted with
$\delta_{\theta}$ and $\delta_{r}$ as in the figure.}
\end{figure}

The nonfactorization of the wavefunction in Eq. (8) indicates the
entanglement of the atom--photon system. In both theoretical
\cite{3-D spontaneous,photoionization} and experimental \cite{exp}
studies of the continuous entanglement, the ratio (denoted by
``$R$'') of the conditional and unconditional variances plays a
central role, since it is a straightforward experimental measure
of the nonseparability, i.e., the entanglement, of the system.

With the single--particle measurement, the unconditional variance
for the effective momentum of the atom is $\delta q^{{\rm
single}}\equiv \langle \Delta q^{2} \rangle-\langle \Delta q
\rangle^{2}$, where the average $\langle\cdot\rangle$ is taken
over the whole ensemble (cf. Eq. (24)). Meanwhile, the coincidence
measurement gives the conditional variance as $\delta q^{{\rm
coin}}\equiv \langle \Delta q^{2} \rangle_{\Delta k_{0}} -\langle
\Delta q \rangle^{2}_{\Delta k_{0}}$, where the photon is now
assumed to be detected at some known $\Delta k_{0}$ (cf. Eq.
(25)). With the two variances, we have:
\begin{eqnarray}
R\equiv\delta q^{{\rm single}}/\delta q^{{\rm coin}}\geq 1.
\end{eqnarray}

Since the $R$ ratio is constructed from the amplitude information
of the wavefunction, it evaluates the ``amplitude entanglement''
for the system; and as it is defined for the momentum
measurements, its value does not vary with time \cite{3-D
spontaneous,phase entang.} and can be directly detected in
experiments \cite{exp}.

\begin{figure}
\centering
\includegraphics[height=4.2cm]{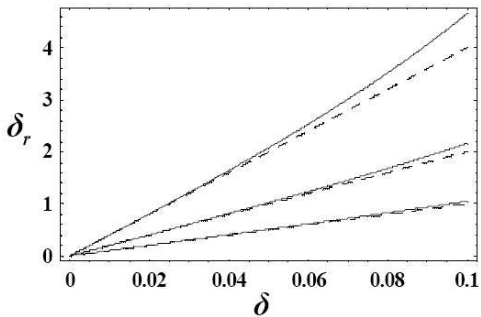}
\caption{The FWHM of the $R$ ratio is plotted in solid lines in
dependence on $\delta$ with $\eta=0.05,0.1,0.2$ from the top to
the bottom. Dashed lines are the fitted function $2\delta/\eta$.}
\end{figure}

Due to the interference with the transitions, the $R$ ratio highly
depends on the initial coherence of the two upper levels, which
can be described with a couple of parameters $(r,\theta)$ defined
as $e^{r+i\theta}=A_{10}/A_{20}$, where $r$ controls the relative
occupation probabilities for the two upper levels, and $\theta$
determines their coherent phase. In further discussions, we assume
$\gamma_{1}=\gamma_{2}\equiv \gamma$ for simplicity, and define a
dimensionless small parameter $\delta\equiv \omega_{12}/\gamma<1$
since the upper levels are nearly degenerate.

With Eqs. (8) and (14), we get the relations between $R$ and the
coherence parameters $r$ and $\theta$ as in Fig. 2. Under the
conditions $\eta\ll 1$ and $\delta^{2}/\eta \ll 1$
\cite{explanations}, we find that the function $R(r,\theta)$ can
be well approximated by a Lorentzian shape, and as shown in Fig. 2
(b), the parameters $r$ and $\theta$ play very symmetric roles in
controlling the ``amplitude entanglement'' $R$, although they are
defined with quite different physical essence.

With the above approximations, we find that the $R$ ratio is
maximized at the ``dark state coherence'', i.e.,
\begin{eqnarray}
R_{{\rm max}}=R(r=0,
\theta=\pi)\approx\frac{\sqrt{2\pi}\eta}{\delta^{2}},
\end{eqnarray}
whereas $r=0$ and $\theta=0$ minimizes the value of $R$.
Furthermore, the full width at half maximum (FWHM) of the function
$R(r,\theta)$ can be well approximated by:
\begin{eqnarray}
\delta_{r}\approx\delta_{\theta}\approx\frac{2\delta}{\eta},
\end{eqnarray}
as shown in Fig. 3. Therefore, with properly chosen atomic
parameters $\eta$ and $\delta$, this scheme could be used to
produce significant ``amplitude entanglement'' in a relatively
large range of the atomic coherence. For example, with
$\eta=\delta=0.01$, the ``amplitude entanglement'' of $R>100$ can
be produced within the range of $0.018<|A_{10}/A_{20}|^{2}<55$,
and $0.36\pi <\theta<1.6\pi$.

\section{full entanglement and steady phase entanglement in momentum}

\begin{figure}
\centering
\includegraphics[height=4cm]{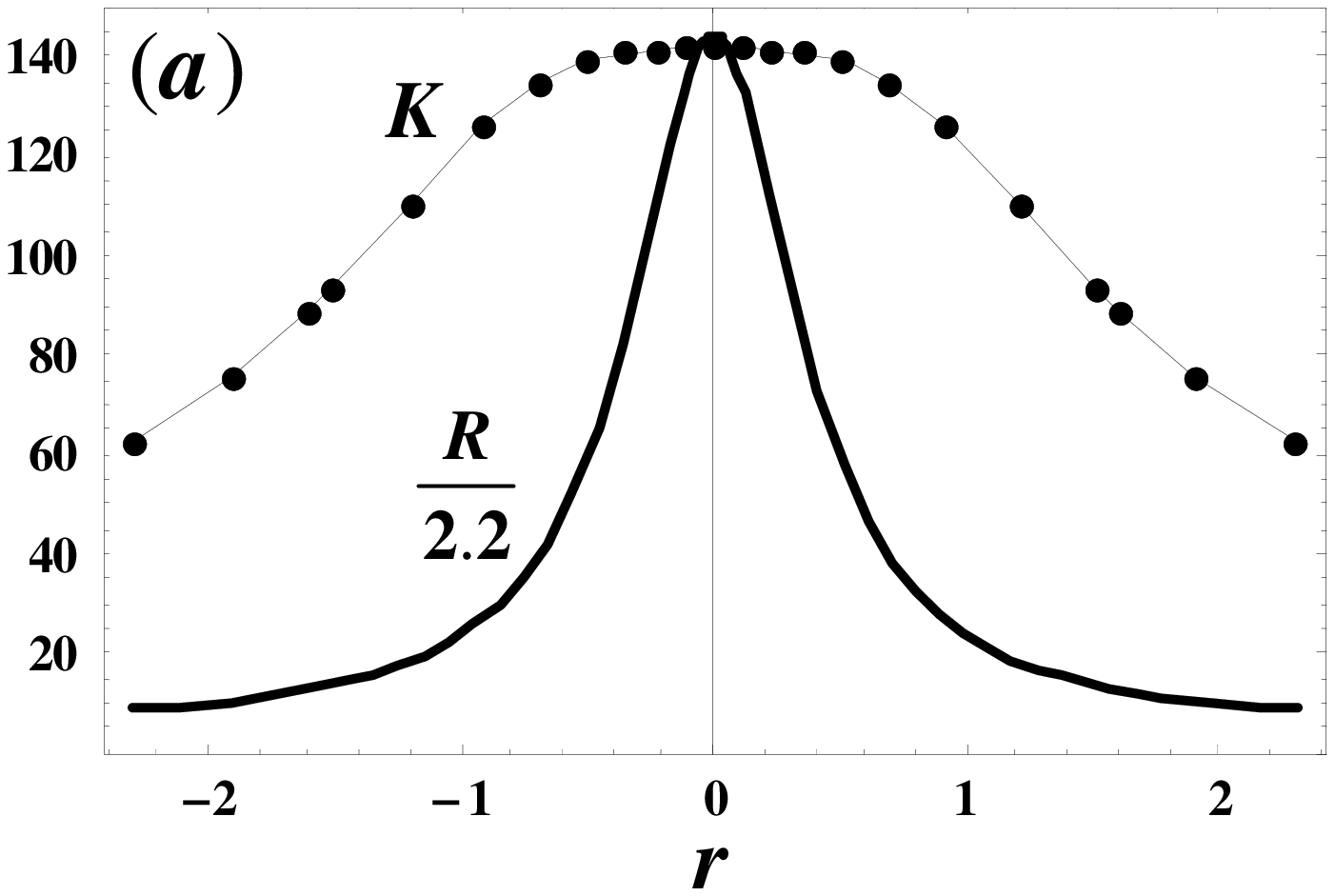}
\includegraphics[height=4cm]{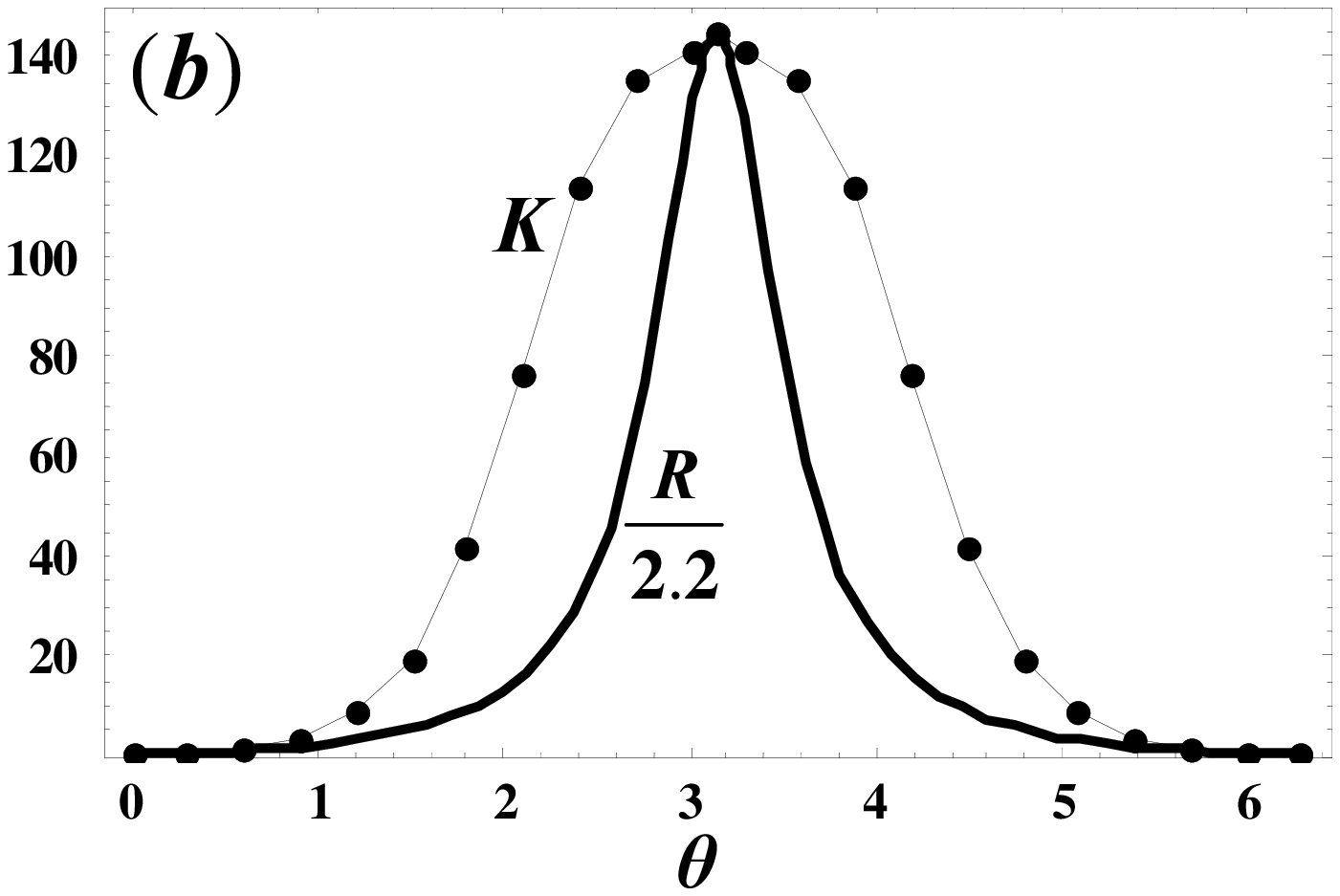}
\includegraphics[height=4cm]{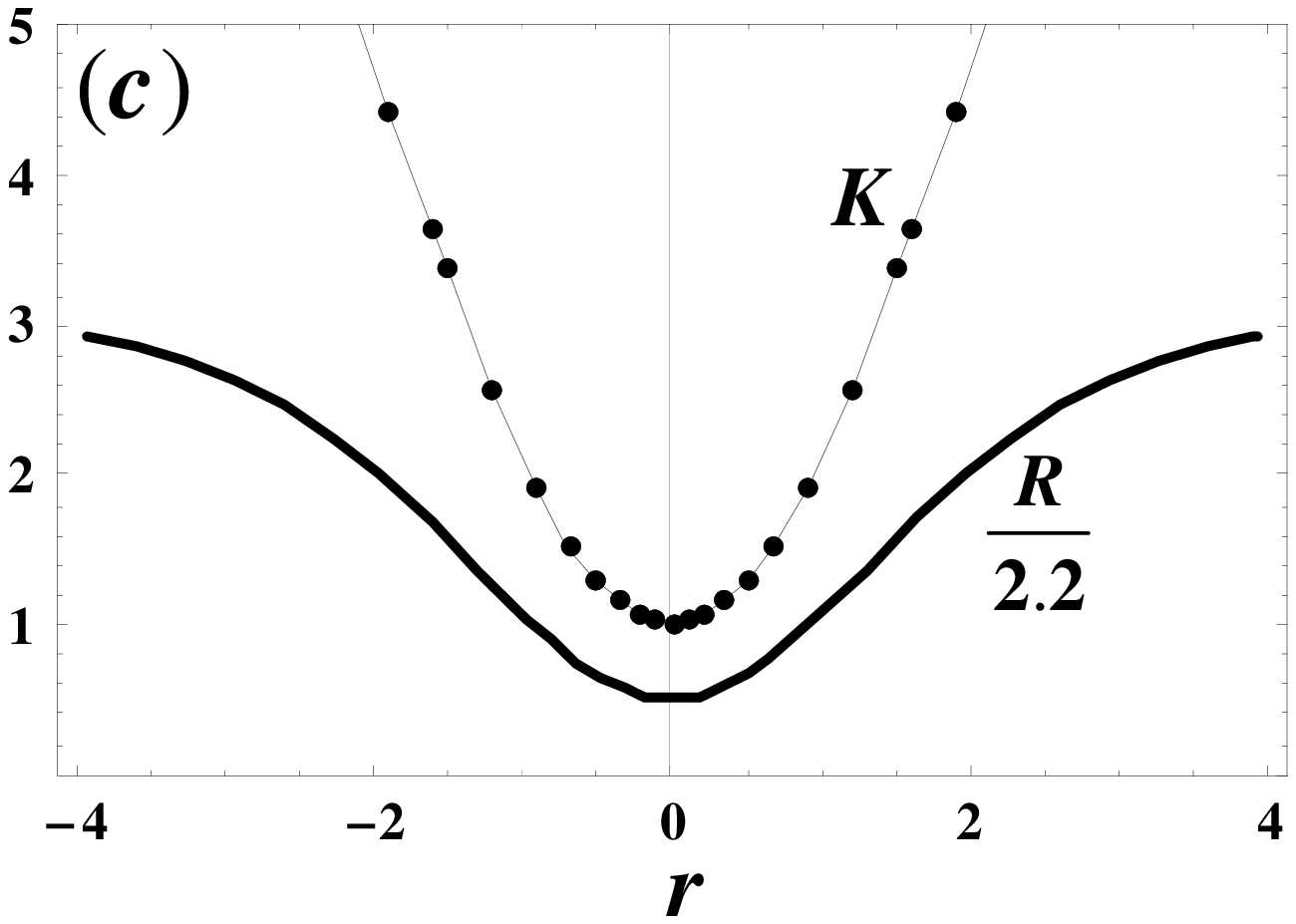}
\caption{Plots of $K$ and $R/2.2$ in dependence on the atomic
coherence $r$ or $\theta$. (a) $\eta=0.2$, $\delta=0.04$, $\theta$
is fixed at $\pi$. (b) $\eta=0.2$, $\delta=0.04$, $r$ is fixed at
$0$. (c) $\eta=0.5$, $\delta=0.08$, $\theta$ is fixed at $0$. }
\end{figure}

In order to evaluate the full entanglement for the bipartite
system in a pure state, one may use the ``Schmidt
number''\cite{Singlephoton,Parametric Down Conversion}. With the
method of Schmidt decomposition \cite{Schmidt num}, the entangled
wavefunction can be uniquely converted into a discrete sum as:
\begin{eqnarray}
B(q,k)=\sum_{n}\sqrt{\lambda_{n}}\psi_{n}(q)\phi_{n}(k),
\end{eqnarray}
where the $\lambda_{n}$'s are ordered as $\lambda_{1}\geq
\lambda_{2}\geq \lambda_{3}\geq ...$ and the $\psi_{n}(q)$'s and
$\phi_{n}(k)$'s are complete orthonormal sets for the Hilbert
spaces of the atom and the photon, respectively. With Eq. (17),
the Schmidt number $K$ is defined as:
\begin{eqnarray}
K\equiv\frac{1}{\sum_{n}\lambda_{n}^{2}}\geq 1.
\end{eqnarray}
As the Schmidt number is constructed with full information of the
wavefunction, and is invariant under representation
transformations, it represents the full entanglement information
for the entangled system.

We plot the numerical results of $K$ in comparison with the $R$
ratio in Fig. 4, where one sees that, both of them are maximized
at the coherence of $(r=0,\theta=\pi)$ and minimized at
$(r=0,\theta=0)$. However, compared with $R(r,\theta)$,
$K(r,\theta)$ has a much slower decay around its maximum, which
indicates that, by tuning the atomic coherence, more entanglement
may be transferred into the phase and can no longer be observed by
the amplitude detection with the $R$ ratio. Actually, due to the
``phase entanglement'', the systems with the same Schmidt number
may exhibit significantly different ``amplitude entanglement''
under different initial conditions. For example, with parameters
$\delta=0.04, \eta=0.7, r=0, \theta=\pi$, for the system we have
$K\approx490$ and $R\approx 1200$; however, when the initial
conditions change to $\eta'=1, r'=0.4$, the system has the same
Schmidt number $K'= K\approx 490$ but a significantly smaller
``amplitude entanglement'' $R'\approx 96 \approx 0.08 R$, because
more entanglement information is transferred into the phase.

Similar phenomenon of the ``phase entanglement'' has been reported
recently \cite{3-D spontaneous,photoionization,phase entang.} in
the position space.  Due to the spreading of the wavepacket, the
phase entanglement in position space appears only in a short time
interval and must be detected by a series of spatial measurements
in time \cite{photoionization}. However, in this scheme, as the
phase entanglement is in the momentum space, it is not affected by
the wavepacket's spreading and keeps invariant with time,
therefore, it will be much easier to be observed in experiments
with direct detections \cite{exp}.

\begin{figure}
\centering
\includegraphics[height=4.8cm]{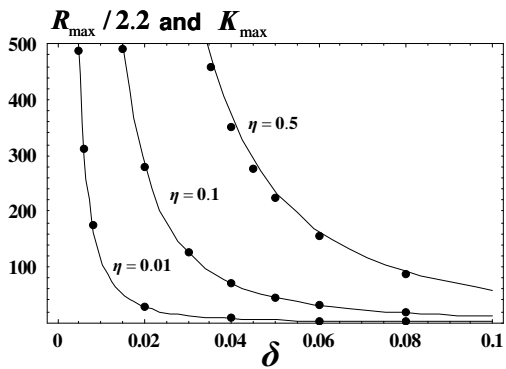}
\caption{The numerical results of $K$ (in spots) and the function
$R/2.2$ (in solid line) are plotted at the maximum atomic
coherence $r=0$ and $\theta=\pi$.}
\end{figure}

By comparing with the ``maximally entangled states'', it is
possible to formally evaluate the ``phase entanglement'' for all
the states under this scheme. For the states with entanglement
maximized by $r$ and $\theta$, the wavefunctions take the form:
\begin{eqnarray}
B(q,k,t\rightarrow\infty)\approx\frac{\chi_{0}\cdot e^{-(\Delta
q/\eta)^{2}}}{i(\Delta q+\Delta k )-\delta^{2}/4}\ ,
\end{eqnarray}
and then the Schmidt number can be well approximated as
\cite{Singlephoton,scattering}:
\begin{eqnarray}
K_{{\rm max}}\approx 1+0.28(4\eta/\delta^{2}-1).
\end{eqnarray}
With Eqs. (20) and (15), we have:
\begin{eqnarray}
K_{{\rm max}}&&=K(r=0,\theta=\pi)\approx\frac{R_{{\rm
max}}}{2.2}\ ,\\
&&\approx \frac{1.12\hbar k_{0}\delta q\gamma}{m \omega_{12}^{2}}.
\end{eqnarray}
As shown in Fig. 5, the relation of Eq. (21) is well fulfilled for
all $\eta$ and $\delta$ with $\eta/\delta^{2}\gg1$ and $\eta\ll 1$
\cite{explanations}.

The linear relation between $K_{{\rm max}}$ and $R_{{\rm max}}$ in
Eq. (21) indicates that there is little phase entanglement for the
``maximally entangled states'' produced with $r=0$ and
$\theta=\pi$, since their full entanglement can be completely
obtained by the amplitude detection with the $R$ ratio. For states
with other $r$ and $\theta$, as in Fig. 4, we always have $K\geq
R/2.2$, therefore, the phase entanglement can be evaluated by:
\begin{eqnarray}
PE\equiv \frac{2.2K}{R}\geq 1.
\end{eqnarray}
The quantity $PE$, as we stated above, evaluates the degree of the
``phase entanglement'' for all the states produced with the
control parameter $\eta$, $\delta$, $r$ and $\theta$ under this
scheme.

In the previous works on atom--photon momentum entanglement
\cite{Singlephoton,3-D spontaneous,scattering,GR}, the
entanglement is produced along a single quantum pathway without
interference. Therefore, the system has a similar wavefunction as
Eq. (19) and exhibits little ``phase entanglement'' in momentum as
we explained above. In this scheme, the interference between two
quantum pathways produce obviously different entangled state as in
Eq. (8), which may give rise to the significant ``phase
entanglement'' in the momentum space.

Moreover, for the single--path scheme \cite{Singlephoton,3-D
spontaneous,scattering,GR}, the Schmidt number is always inversely
proportional to the linewidth of the transition; while in our
proposed scheme, however, one sees that $K_{{\rm max}}\propto
\gamma$ as in Eq. (22). This abnormal phenomenon indicates that
the mechanism for the entanglement in this scheme is essentially
different with the previous ones \cite{Singlephoton,3-D
spontaneous,scattering,GR}. By squeezing the effective separation
of the upper levels, as $K\propto 1/\omega_{12}^{2}$ in Eq. (22),
it is possible to use this scheme to produce supper--high degree
of momentum entanglement for the atom--photon systems
\cite{GR-disentanglement}.

\section{Schmidt modes}

\begin{figure}
\centering
\includegraphics[height=5.8cm]{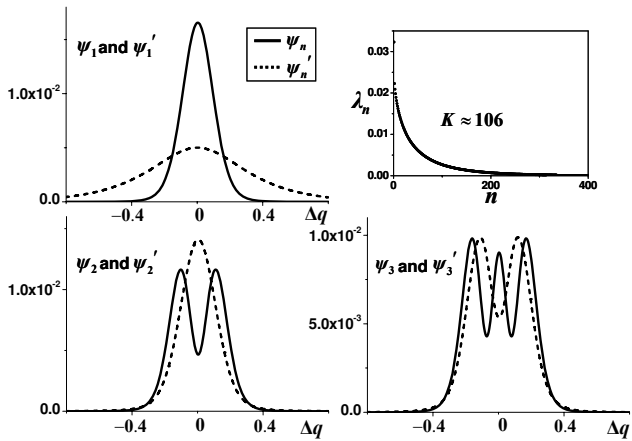}
\caption{First three Schmidt modes are compared between states
$B(q,k)$ and $B'(q,k)$ with $\delta=0.1$, $\eta=0.94$, $r=0$,
$\theta=\pi $ and $\delta'=\delta$, $\theta'=\theta$, $\eta'=1$,
$r'=-0.4$, respectively, where one has $K'\approx K$ and
$R'\approx 0.3R$. The phase entanglement of $B'(q,k)$ broadens the
fist Schmidt mode and decreases the number of peaks for the rest
ones. The inset shows the distribution of the eigenvalues
$\lambda_{n}$ in the Schmidt decomposition.}
\end{figure}

The phenomenon of ``phase entanglement'' is related to the
coherence between different Schmidt modes $\psi_{n}(q)$ defined in
Eq. (17). With the Schmidt decomposition, the unconditional
variance $\delta q^{{\rm single}}$ can be written as:
\begin{eqnarray}
\delta q^{{\rm single}}&=&\int {\rm d} \Delta q\ {\rm d} \Delta k
\Delta q^{2} |B(q,k)|^{2}, \nonumber  \\
&=& \int {\rm d} \Delta q\ \Delta q^{2}\sum_{n}\lambda_{n}|\psi_{n}(q)|^{2} , \nonumber \\
&=& \int {\rm d} \Delta q\ \Delta q^{2}E_{{\rm i}}(q),
\end{eqnarray}
where $E_{{\rm i}}(q)\equiv \sum_{n}\lambda_{n}|\psi_{n}(q)|^{2}$
is a probability distribution constructed by an ``incoherent''
summation of different Schmidt modes weighed by $\lambda_{n}$. The
conditional variance, however, is written as:
\begin{eqnarray}
&& \delta q^{{\rm coin}}= \frac{1}{N}\int {\rm d} \Delta q\ \Delta
q^{2} |B(q,\Delta k_{0})|^{2},\nonumber \\
&&= \frac{1}{N} \int {\rm d} \Delta q\ \Delta
q^{2}|\sum_{n}\sqrt{\lambda_{n}}\phi_{n}(\Delta
k_{0})\psi_{n}(q)|^{2},
\end{eqnarray}
where $N=\sum_{n}\lambda_{n}|\phi_{n}(\Delta k_{0})|^{2}$ is the
normalized coefficient. When $K$ is large, $\phi_{n}(\Delta
k_{0})$ can be approximated as a constant, then we have:
\begin{eqnarray}
\delta q^{{\rm coin}}\approx \int {\rm d} \Delta q\ \Delta q^{2}
|E_{{\rm c}}(q)|^{2},
\end{eqnarray}
where $E_{{\rm c}}(q)\equiv \sum_{n}\sqrt{\lambda_{n}}\psi_{n}(q)$
is a ``coherent superposition'' of different Schmidt modes.
Comparing Eq. (24) with (26), one sees that the $R$ ratio defined
as $R=\delta q^{{\rm single}}/\delta q^{{\rm coin}}$ actually
represents the degree of the packet narrowing caused by the
coherence between different Schmidt modes.

In Fig. 6, we compare the atomic Schmidt modes between two states
$B(q,k)$ and $B'(q,k)$ with $K'\approx K$ and $R'\approx 0.3 R$.
It is found that the phase entanglement will significantly broaden
the fist few Schmidt modes and decrease the number of peaks for
the rest ones. Moreover, the coherence between different Schmidt
modes diminishes, which decreases the $R$ ratio as we stated
above.

The photonic Schmidt modes exhibit similar properties as the
atomic modes. We emphasize that the property of Gaussian
localization \cite{Singlephoton,scattering} of the single--photon
modes still remains in spite of the shape distortions caused by
the interference. Therefore, it demonstrates the possibilty to
apply the idea of ``localized single--photon wavefunction in free
space''\cite{Singlephoton} to even more complicated atomic
systems.

\section{conclusion}

In summary, we have investigated the recoiled--induced
atom--photon entanglement in the atomic system with SGC. Due to
the quantum interference in the emission process, the momentum
entanglement can be effectively controlled by the atomic internal
coherence, and may be greatly enhanced by increasing the linewidth
or squeezing the separation of the upper levels as in Eq. (22).
The novel phenomenon of ``momentum phase entanglement'' is shown
and evaluated quantitively. Further, we compare the atomic Schmidt
modes for different entangled states in the momentum space.

In order to experimentally observe these phenomena, one needs two
nearly degenerate upper levels with parallel dipole moments. This
configuration has been extensively studied both theoretically and
experimentally \cite{SGC,atomic coherence contr} in recent years,
and can be realized by mixing different parity levels or by using
dressed-state ideas. With proper control of the atomic coherence
\cite{atomic coherence contr}, it is most probable to observe the
``momentum phase entanglement'' in experiments. Furthermore, by
squeezing the separation of the upper-levels in dressed--state
with an auxiliary light \cite{SGC}, this scheme can be used to
produce super-high degree of entanglement for realistic applications \cite{GR-disentanglement} .\\

This work is supported by the National Natural Science Foundation
of China (Grant No. 10474004), and DAAD exchange program:
D/05/06972 Projektbezogener Personenaustausch mit China
(Germany/China Joint Research Program).


\begin{references}
\bibitem{explanations} In typical atomic system, $\eta$ is of order $0.1$ or smaller, e.g., for sodium one has
$\eta\approx 0.05$; moreover, the condition $\delta^{2}/\eta\ll 1$
is equivalent to $K_{{\rm max}}\gg 1$, which is our major concern.
\end{references}
\end{document}